\documentclass[epj]{svjour}
\usepackage{graphics}
\usepackage{amssymb}
\usepackage{delarray,amsmath}
\usepackage[table]{xcolor}
\begin{document}
\title{Simplest phonons and pseudo-phonons in field theory}
\author{Daniele Musso
\thanks{daniele.musso@usc.es}%
}                     
%
%
\institute{Universidad de Santiago de Compostela (USC) \and Instituto Galego de F\'isica de Altas Enerx\'ias (IGFAE)}
\date{\phantom{Received: date / Revised version: date}}
%
\abstract{
We study the emergence of Nambu-Goldstone modes due to broken translation symmetry in field theory.
Purely spontaneous breaking yields a massless phonon which develops a mass upon introducing
a perturbative explicit breaking. 
The pseudo-phonon mass agrees with Gell Mann-Oakes-Renner relations. 
We analyze the simplest possible theories featuring gradient Mexican hats
and describing space-dependent order parameters.
We comment on homogeneous translation breaking and the connections
with holographic Q-lattices.
%
} 

\maketitle
\section{Introduction and motivation}

The breaking of spatial translations is fundamental to many physical systems,
especially but not only in condensed matter. 
The lack of momentum conservation and the emergence 
of phonons are two important consequences of different kinds of translation symmetry breaking,
 \emph{explicit} and \emph{spontaneous} respectively. 
Typically the realization of translation symmetry breaking is 
technically complicated due to the spatial dependence of the fields.
Although not being a conceptual obstruction, 
such complication usually requires non-analytic tools. 

An important exception to this state of affairs 
has been explored in recent years within the holographic context, 
with the phenomenological aim of modeling strongly-correlated systems that break translations
and lack a standard quasi-particle description. 
Holographic models geometrize the renormalization group flow 
of a conjectured quantum field theory by means of dual gravitational fields, 
often having a radial dependence 
in an otherwise symmetric and homogeneous space. 
Spatial features can be added, 
yet the computations get much more difficult 
since the equations of motion become partial differential equations.

Wide families of holographic models avoid the leap in complexity at the price of introducing conceptual subtleties.
Such models are characterized by a \emph{homogeneous} breaking of translation symmetry, where an internal global
symmetry renders all the spacetime points of the broken phase equivalent \cite{Donos:2013eha,Andrade:2013gsa,Donos:2014oha}.%
\footnote{Models featuring massive gravity \cite{Vegh:2013sk,Davison:2013jba,Amoretti:2014zha} have been considered too, 
they fall into the same category because they can be realized through a Stueckelberg mechanism in terms of fields that enjoy a global symmetry \cite{deRham:2011rn,Baggioli:2014roa}.
Massive gravity has also been employed to define effective field theories for phonons \cite{Alberte:2017oqx,Alberte:2018doe}.}
Basic concepts like those of \emph{unit cell} or \emph{commensurability} \cite{Andrade:2015iyf}
are thereby absent.

The present paper exports the idea of homogeneous translation symmetry breaking to standard, non-holographic field theory.
Specifically, it studies the ``Mexican hat potential'' for translation symmetry breaking, both in a literal and a loose sense:
literally, the mechanism of the Mexican hat is realized in the gradient sector; 
loosely, the aim at stake is to analyze simple prototypical field theory examples of
translation symmetry breaking.

Phonons have been long considered in field theory, yet they are introduced \emph{ad hoc}.
This is the case for Fr\"ohlich Hamiltonians describing electron-phonon interactions,
and for effective field theories of elastic media \cite{Leutwyler:1996er}. 
Many phenomenological questions indeed are insensitive to the microscopic details. 
Nevertheless, there are reasons why elucidating the origin of phonons 
is important, both conceptually and phenomenologically:%
\footnote{The present paper describes phonons corresponding to \emph{sliding modes} of spatially modulated order parameters,
not to be confused with the phonons due to the ionic lattice.}.
\begin{itemize}
\item Low-energy effective field theories relies on information about the symmetries:
the number of low-energy degrees of freedom is in general expected to be given by counting theorems 
descending from the symmetry breaking pattern \cite{Low:2001bw}.
\item Spontaneous pattern formation is at the basis of the density waves physics \cite{Gruner:1988zz,interco}.
Dynamical density waves are the best candidates to explain the anomalous transport properties
of many strongly-correlated electron systems (\emph{e.g.} high-$T_c$ superconductors both 
in the ``normal'' and ``condensed'' phases \cite{interco}).%
\item Implementation and check of general expectation about the consequences of the symmetry breaking.
\end{itemize}
In this latter sense, we prove the validity of Gell Mann-Oakes-Renner relations for pseudo-phonons:%
\footnote{The word \emph{pseudo} refers to the spontaneous breaking of an approximate symmetry.} 
their squared mass is, at leading order, linear in the perturbation which breaks translations explicitly.

A perturbative explicit component of translation symmetry breaking can model the effects of weak disorder.
In particular, it \emph{pins} a modulated order parameter preventing it from sliding freely.
Weakly pinned density waves are the best candidate mechanism to explain the \emph{bad metal} phenomenology \cite{MIR,Delacretaz:2016ivq,Delacretaz:2017zxd}.
The argument goes as follows: despite having little disorder,
bad metals have exceptionally low dc conductivity due to a significant spectral weight transfer to higher frequencies. 
In other words, the optical conductivity has a marked peak at finite frequency, 
which subtracts spectral weight from null frequency. 
The soft finite-frequency peak can have a direct connection to a pseudo-phonon associated to a 
weakly-pinned, modulated order parameter. 
This hypothesis is further corroborated by the ubiquitous presence 
of spatial patterns throughout the phase diagram of strongly-correlated electron systems \cite{PhysRevB.54.7489,2018NatMa,2018arXiv180904949A}.

The models studied in the present paper shed light on possible field theory duals to holographic models 
which break translation symmetry homogeneously.%
\footnote{The models analyzed in this paper have indeed been suggested by holographic theories, see for instance 
\cite{Donos:2013eha,Amoretti:2016bxs,Amoretti:2017frz,Amoretti:2017axe}. In a wider historical perspective,
it is fair to acknowledge the Q-ball construction as an inspiring progenitor \cite{Coleman:1985ki}.}

\section{Main results}
The main results of the paper are:
\begin{enumerate}
 \item The emergence of phonons in a generic class of field theories \eqref{GM},
 and the characterization of their dispersion relation \eqref{dispe_massless}.
 \item The addition of a perturbative term that breaks translation symmetry explicitly 
 and the appearance of a mass for the phonon \eqref{pse}.
 \item Realization of homogeneous translation symmetry breaking in a purely field theoretic model (Section \ref{homo}).
 \item Construction of toy-models for the formation of concomitant density waves at an angle (Section \ref{WC}).
\end{enumerate}

\section{Phonon and pseudo-phonon}
\label{CDWlike}

\subsection{Spontaneous breaking of translations}

Consider the following action
\begin{equation}\label{GM}
 \begin{split}
 S &= \int d^3x \ \Big\{
 -(\partial^t \phi^*)(\partial_t \phi) 
 +A (\partial^i \phi^*)(\partial_i \phi)\\ &
 -B \left[(\partial^i \phi^*)(\partial_i \phi)\right]^2
 -F\, \phi^*\phi\, (\partial^i \partial^j \phi^*)(\partial_i \partial_j \phi)\\ &
 +G\, (\partial^i \phi^*)(\partial_i \phi^*)(\partial^j \phi)(\partial_j \phi)
 \Big\}\ ,
 \end{split}
\end{equation}
for a scalar complex field $\phi$ where $A$, $B$, $F$ and $G$ are real positive numbers and $i$, $j$ are spatial indexes, $i,j=1,2$.
The metric signature is $(-1,1,1)$.
The terms controlled by the couplings $B$, $F$ and $G$ are fourth-order both in the spatial derivatives and in the field.
It is convenient to parametrize the complex field in terms of the modulus and the phase fields
\begin{equation}\label{polar}
 \phi(t,x,y) = \rho(t,x,y)\, e^{i \varphi(t,x,y)}\ .
\end{equation}

Model \eqref{GM} is spatially isotropic but breaks Lorentz invariance explicitly,
it enjoys spacetime translation invariance and global $U(1)$ phase rotations of $\phi$ (namely shift symmetry for the phase field $\varphi$).
The Euler-Lagrange equations for $\phi$ are given by
\begin{equation}\label{EOMgen}
 \begin{split}
 \text{EOM}[\phi] &=
\partial^t\partial_t \phi
-A \partial^i\partial_i \phi
+2B \partial^i \left[\partial^j\phi^*\partial_j\phi\, \partial_i\phi\right]\\
&-F\, \partial^i \partial^j[\phi^*\phi\, \partial_i\partial_j \phi]
-F\, \phi\, (\partial^i \partial^j \phi^*)(\partial_i \partial_j \phi)\\
&-2G\, \partial^i [\partial_i \phi^*\partial^j \phi\partial_j \phi]
=0\ ,
 \end{split}
\end{equation}
and its complex conjugate equation $\text{EOM}[\phi^*]$.
Recall that the variational problem leading to the equations of motion assumes $\delta \phi = \delta \phi^* = 0$ at infinity.
Passing to the ``polar representation'' \eqref{polar},
one can obtain the equations of motion for the modulus and phase fields 
by means of the combinations
\begin{align}\label{EOMrho}
 \rho\, \text{EOM}[\rho] = \phi^*\, \text{EOM}[\phi] + \phi\, \text{EOM}[\phi^*]\ , \\ \label{EOMvarphi}
 i\, \text{EOM}[\varphi] = \phi^*\, \text{EOM}[\phi] - \phi\, \text{EOM}[\phi^*]\ .
\end{align}

Consider the following static but $x$-dependent ansatz:
\begin{equation}\label{ansa}
 \rho(t,x,y) = \bar \rho\ , \qquad
 \varphi(t,x,y) = k\, x\ ,
\end{equation}
and plug it into the modulus equation of motion \eqref{EOMrho}, thus obtaining
\begin{align}\label{onEOM}
  \text{EOM}[\rho] = 2 k^2 \bar\rho^2\ \left[A - 2 k^2 \bar\rho^2 \left(B+F-G\right)\right] = 0\ ,
\end{align}
while the equation of motion for $\varphi$ \eqref{EOMvarphi} is automatically satisfied. 
The solutions to \eqref{onEOM} are%
\footnote{We are solving $k$ as a function of $\rho$, regarding the latter as fixed 
to a non-trivial value by boundary conditions. Comments on the handling 
of boundary conditions are given later, especially in Section \ref{pota}.}
\begin{align}\label{sole1}
 k^{(1,2)} &= 0\ ,\\ \label{sole2}
 k^{(3,4)} &= \pm\, \frac{1}{\bar\rho}\left[\frac{A}{2(B+F-G)}\right]^{\frac{1}{2}}\ .
\end{align}
To have real solutions we demand $B+F >G$.
\begin{figure}
\resizebox{0.47\textwidth}{!}{%
  \includegraphics{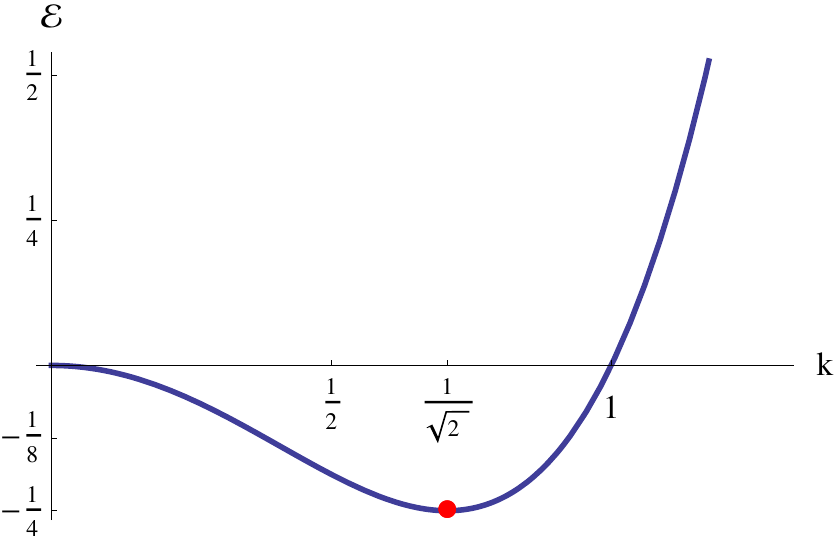} 
}
\caption{Plot of the static energy for the particular case $A=B=F=G=\bar\rho=1$.
The red dot corresponds to the global minimum of the energy, $-1/4$, 
 attained for $k=1/\sqrt{2}$.}
\label{ene}       
\end{figure}
The energy density of a static configuration of the form \eqref{ansa} is given by
\begin{equation}\label{enel}
 {\cal E}(k,\bar\rho) = k^2 \bar\rho^2 \left[-A + k^2 \bar\rho^2 \left(B+F-G\right) \right]\ ,
\end{equation}
and it is spatially homogeneous.
In particular, on the solutions \eqref{sole1} and \eqref{sole2} the energy attains 
the following values 
\begin{align}
 {\cal E}(k^{(1,2)}) &= 0 \ , \\ \label{ene2}
 {\cal E}(k^{(3,4)}) &= - \frac{1}{2} A (k^{(3,4)})^2 \bar\rho^2 \ .
\end{align}
Notice that ${\cal E}(k^{(3,4)})$ represent degenerate global minima because $A>0$.
In Figure \ref{ene} we plot a specific example.

Having specified the background of interest \eqref{ansa}, let us consider now the fluctuations 
\begin{equation}
 \phi(t,x,y) = \Big[\bar \rho + \delta\eta(t,x,y)\Big] e^{i kx}\ ,
\end{equation}
and parametrize the complex fluctuation field $\delta\eta$ by its real and imaginary parts
\begin{equation}\label{flut}
 \delta \eta(t,x,y) = \sigma(t,x,y) + i \tau(t,x,y)\ .
\end{equation}
The quadratic action for the fluctuations in Fourier space is given by
\begin{equation}\label{S2}
 S_{(2)} = \int \frac{d^3 q}{(2\pi)^3} \ \tilde v(-q)^T \cdot M \cdot \tilde v(q)\ ,
\end{equation}
where $q=(\omega,q_x,q_y)$ and
\begin{equation}
 \tilde v(q) = \left(\begin{array}{c} \tilde \sigma(q) \\ \tilde \tau(q) \end{array}\right)\ .
\end{equation}
The entries of the matrix $M$ defined in \eqref{S2} are given by
\begin{align}
 M_{\sigma\sigma} &= \omega^2 - 2 A k^2 - 4 k^2 \bar\rho^2 (2Fq_x^2 + G q_y^2) \\ \nonumber &\qquad - F \bar\rho^2 (q_x^2+q_y^2)^2\ ,\\ \label{tautau_spo}
 M_{\tau\tau} &= \omega^2-2 A q_x^2 - F \bar\rho ^2 \left(q_x^2+q_y^2\right)^2\ ,\\ 
 M_{\sigma\tau} &= M_{\tau\sigma}^* = -2 i k\, q_x \left[A + 2 F \bar\rho ^2\left(q_x^2+q_y^2\right)\right]\ .
\end{align}
The eigenvalues $e_{1,2}$ of $M$ are given by complicated expressions.
To highlight the physical characteristics of the two modes,
it is convenient to expand the dispersion relations $e_{1,2}=0$ for low momenta, thus obtaining
\begin{align}\label{dispe_massive}
\omega^2 =&\ 2 A k^2 
+ 4 k^2 \bar \rho^2 [(B+3F-G) q_x^2 + G q_y^2]
+ ...
\ ,
\end{align}
where $k$ is given by \eqref{sole2}, and
\begin{align}\label{dispe_massless}
\omega^2 = \bar\rho^2 \left[2(2 G-3 F)q_x^2 q_y^2 + F(q_x^4+ q_y^4)\right] 
+ ... \ .
\end{align}
In \eqref{dispe_massless}, the positivity of the quartic term in momenta requires $G\geq F$.
We also require $F\neq0$. Indeed, for $F=0$ there would be neither pure longitudinal ($q_y=0$)
nor pure transverse ($q_x=0$) propagation in \eqref{dispe_massless}.
Note that $F\neq0$ implies that we are keeping in the action \eqref{GM}
a term with more than one spatial derivative acting on a single field. 
Since a similar term in time derivatives would lead to Ostrogradsky instabilities \cite{Ostrogradsky:1850fid,Woodard:2015zca}, 
the condition $F\neq0$ constitutes an obstruction to building simple relativistic generalizations of model \eqref{GM}.
Combining the requirements above we have
\begin{equation}\label{costra}
 G\geq F>0\ .
\end{equation}

Equation \eqref{dispe_massless} describes a gapless mode which is the \emph{phonon}.%
\footnote{We use the term \emph{phonon} to indicate the Nambu-Goldstone mode arising 
from the spontaneous breaking of spatial translations.}
It represents the Nambu-Goldstone mode arising from the spontaneous breaking
of the product of phase shifts and translations to its diagonal subgroup.
The phonon dispersion relation is not standard, because $\omega$ is not linear in the momentum%
\footnote{It is interesting to note that a quadratic dispersion relation was obtained for the 
scalar mode of the ghost condensate \cite{ArkaniHamed:2003uy} too. 
In the literature one can encounter other examples of phonons with a quadratic dispersion relation \cite{2016arXiv160102884C}.}, 
and the propagation is in general not isotropic (despite being $x\leftrightarrow y$ symmetric).
The dispersion relation given in \eqref{dispe_massive} corresponds instead to a gapped mode. 
Both its mass and the leading $q_x$ term are not essentially related to the 
couplings $F$ and $G$, but the transverse propagation at quadratic order would vanish for $G=0$. 
In general such gapped mode does not propagate 
isotropically and its dispersion relation is not symmetric under $x\leftrightarrow y$ 
exchange.

As a final comment, studying the fluctuations about a background
\eqref{sole1} with $k=0$, one finds that it is locally unstable.

\subsection{Adding a small explicit translation breaking}
\label{expl}

We add to the action \eqref{GM} a perturbation which breaks translations along $x$ explicitly:
\begin{equation}\label{GMp}
 \begin{split}
 S &= \int d^3x \ \Big\{
 -(\partial^t \phi^*)(\partial_t \phi) 
 +A (\partial^i \phi^*)(\partial_i \phi)\\ &
 -B \left[(\partial^i \phi^*)(\partial_i \phi)\right]^2
 -F\, \phi^*\phi\, (\partial^i \partial^j \phi^*)(\partial_i \partial_j \phi)\\ &
 +G\, (\partial^i \phi^*)(\partial_i \phi^*)(\partial^j \phi)(\partial_j \phi)
 +n f_\kappa(x,\phi^*,\phi)
 \Big\}\ ,
 \end{split}
\end{equation}
where $n$ is a perturbative coupling, and $f_\kappa$ is an explicit symmetry-breaking term controlled by the parameter $\kappa$.
Concretely we take:
\begin{equation}\label{exp}
 f_\kappa(x,\phi^*,\phi) = \frac{\kappa^2}{2}\left[\phi\, e^{-i \kappa x} - \phi^*\, e^{i \kappa x}\right]^2\ ,
\end{equation}
which vanishes for $\kappa\rightarrow 0$. 

We consider again the ansatz \eqref{ansa}, this time the equation of motion 
\eqref{EOMvarphi} for the phase field $\varphi$ is not automatically satisfied,
but takes the form
\begin{equation}\label{vara}
 \text{EOM}[\varphi] = - 2i n \kappa^2 \bar\rho^2 \sin[2x(k-\kappa)]\ .
\end{equation}
To solve \eqref{vara} we fix
\begin{equation}\label{kk}
 k = \kappa \ .
\end{equation}
Upon considering \eqref{kk}, the equation of motion for the modulus field $\rho$ reduces again to \eqref{EOMrho}.

The quadratic action for the fluctuations gets modified by the perturbation,
in particular the entries of the matrix $M$ introduced in \eqref{S2} become
\begin{align}
 M_{\sigma\sigma} &= \omega^2 - 2 A k^2 - 4 k^2 \bar\rho^2 (2Fq_x^2 + G q_y^2) \\ \nonumber &\qquad - F \bar\rho^2 (q_x^2+q_y^2)^2\ ,\\ \label{tautau_spo}
 M_{\tau\tau} &= \omega^2 -2 n \kappa^2 -2 A q_x^2 - F \bar\rho ^2 \left(q_x^2+q_y^2\right)^2\ ,\\ 
 M_{\sigma\tau} &= M_{\tau\sigma}^* = -2 i k\, q_x \left[A + 2 F \bar\rho ^2\left(q_x^2+q_y^2\right)\right]\ .
\end{align}
The only difference with respect to the spontaneous case is given 
by a new $-2n\kappa^2$ term in $M_{\tau\tau}$.
Expanding the two dispersion relations $e_{1,2}=0$ in low momenta and in small $n$, one obtains%
\footnote{Recall that $e_{1,2}$ are the eigenvalues of the matrix $M$.}
\begin{align}\label{dispe_massive_n}
\omega^2 = &\ 2 A k^2 + 2 n q_x^2\\ \nonumber
&+ 4 k^2 \bar\rho^2 [(B+3F-G)q_x^2 + G q_y^2] + ...\ ,
\end{align}
and
\begin{align}\label{pse}
&\omega^2 = 2n \kappa^2 - 2 n q_x^2
+\left[F +\frac{4}{A}(B+3F-G)n\right] \bar\rho^2 q_x^4\\ & \nonumber
+2\left[(2G-3F)+\frac{4}{A}(G-F)n\right]\bar\rho^2 q_x^2 q_y^2
+F \bar\rho^2 q_y^4
+...\ ,
\end{align}
where $k$ and $\kappa$ are given by \eqref{kk} and \eqref{sole2}.
The squared mass $2 n \kappa^2$ in \eqref{pse} is linear in the perturbative coupling $n$ which controls the explicit breaking,
this agrees with Gell Mann-Oakes-Renner expectation for a pseudo Nambu-Goldstone mode, 
thereby \eqref{pse} describes a \emph{pseudo-phonon}.%
\footnote{Gell Mann-Oakes-Renner relations for phonons were obtained in holographic models in \cite{Amoretti:2016bxs}.
For an experimental study of gapped sliding modes, see for example \cite{PhysRevB.94.134309}.}
Note also that the explicit breaking affects both the dispersion relations of the gapped and the Nambu-Goldstone 
modes at the quadratic level in $q_x$. This is particularly relevant for the pseudo-phonon 
as it lowers the leading order at which the longitudinal momentum enters the dispersion relation.
As far as the leading transverse propagation is concerned, the explicit breaking term does not alter the 
qualitative picture.

\section{Comments on homogeneous translation symmetry breaking}
\label{homo}

A solution of the form \eqref{ansa} breaks the product of translations along $x$ and $\varphi$-shifts to the diagonal subgroup,
so a translation along $x$ can be compensated by a global phase shift. This is the hallmark of \emph{homogeneous} translation 
symmetry breaking: any spacetime point is equivalent to any other up to a global \emph{internal} transformation. There is 
no \emph{unit cell} and $k$ should not be strictly speaking interpreted as the wave vector of a lattice.%
\footnote{Relatedly, holographic models with backgrounds that break translations homogeneously do not feature \emph{commensurability} effects 
for the formation of stripes \cite{Andrade:2015iyf}.}

The canonical energy-momentum tensor of model \eqref{GM} is given by
 \begin{align}
  &T^{\mu\nu} =
  \frac{1}{2}\eta^{\mu\nu} {\cal L} 
  +\delta^\mu_t\, \partial^t \phi^* \partial^\nu \phi
  +\delta^\mu_i \Big[ A  \partial^i\phi^* \\ \nonumber
  &-2B\partial^j\phi^* \partial_j \phi \partial^i \phi^* 
  +2G\partial^j\phi^*\partial_j\phi^*\partial^i\phi\\ \nonumber 
  &-F\phi^*\phi \partial^i \partial^j\phi^* \partial_j 
  +F\partial_j(\phi^*\phi \partial^i\partial^j\phi^*)\Big] \partial^\nu\phi
  + \text{c.c.}\ .
 \end{align}
The momentum density vanishes $T^{tx}=T^{ty}=0$ on a solution of the form \eqref{ansa}.
On the equations of motion we have $\partial_\mu T^{\mu\nu} = 0$, thus the Ward-Takahashi 
identities for translations are satisfied.

On a solution \eqref{ansa} the energy density is given by
\begin{equation}\label{ene_den}
 \epsilon = T^{tt} = \eta^{tt} L_0 = -L_0\ ,
\end{equation}
where $L_0$ is the Lagrangian density written in \eqref{GM} considered on the background \eqref{ansa}.
Similarly, the spatial components of the energy-momentum tensor are:
\begin{align}\label{Txx} 
 T^{xx} &= \eta^{xx} L_0 + 2 \bar \rho^2 k^2 \left[A - 2\bar \rho^2 k^2 \left(B-G+F\right)\right] = L_0\ , \\ \label{Tyy}
 T^{yy} &= \eta^{yy} L_0 = L_0\ ,
\end{align}
where in the second step of \eqref{Txx} we have used the equation of motion \eqref{onEOM}.
We define the pressure $p= - \Omega/V$ where $\Omega/V$ is the Landau potential density.
Given that $T=\mu=0$, we have $\Omega/V = \epsilon - T s - \mu n = \epsilon$.
Comparing \eqref{ene_den} with \eqref{Txx} and \eqref{Tyy} we obtain $p=T^{xx} = T^{yy}$.
The pressure is thus isotropic and the equation of state is given by%
\footnote{This is the same equation of state found for the ghost condensate \cite{ArkaniHamed:2003uy}.}
\begin{equation}\label{EOS}
 \epsilon = - p\ .
\end{equation}

The $U(1)$ current density is given by
\begin{align}\label{curre}
 J^\mu =&\
 i\delta^\mu_t\partial^t\phi^*\phi
 -i \delta^\mu_i \Big\{A\partial^i\phi^* \phi
 -2B\partial^i\phi^*\partial^j\phi^*\partial_j\phi\phi \\ \nonumber
 &+2G\partial^i\phi\partial^j\phi^* \partial_j\phi^*\phi 
 - F \phi^*\phi\partial^i\partial^j \phi^*\partial_j\phi\\ \nonumber &
+F\partial_j[\phi^*\phi\partial^i\partial^j \phi^*] \phi  \Big\} + \text{c.c.} \ ,
\end{align}
on a solution of the form \eqref{ansa} it vanishes too.

Remarkably, the explicit breaking term \eqref{exp} does not 
introduce a source into the translation 1-point Ward-Takahashi identity.
Indeed, consider the $\phi$ field transformation under a diffeomorphism
\begin{equation}
 \delta_\xi \phi = \xi^\mu \partial_\mu \phi\ ,
\end{equation}
and take $\xi^\mu = \delta^\mu_x \xi$. One has that the explicit breaking term in \eqref{GMp} transforms as follows:
\begin{align}
 & \delta_\xi f_\kappa (x,\phi^*,\phi) = \frac{\delta f_\kappa}{\delta \phi} \delta_\xi \phi + \frac{\delta f_\kappa}{\delta \phi^*} \delta_\xi \phi^* \\ \nonumber
 &= \kappa^2\, \xi \left[\phi e^{-i\kappa x} - \phi^* e^{i\kappa x} \right] \left[e^{-i\kappa x} \partial_x \phi - e^{i\kappa x} \partial_x \phi^*\right]\\ \nonumber
 &= 2 i \kappa^3\, \xi\, \bar \rho \left(\bar \rho - \bar \rho \right) = 0\ ,
\end{align}
where in the last passages we have used both the ansatz \eqref{ansa} and the condition \eqref{kk}.
The triviality of the 1-point Ward-Takahashi identity for translations, in spite of the presence of 
a term which breaks translations explicitly, is a peculiarity of homogeneous breakings.
These have already been studied in holographic models,
see \cite{Amoretti:2016bxs} for instance.

Analogous arguments show that the explicit term \eqref{exp} 
does not introduce a source into the Ward-Takahashi identity 
of the $U(1)$ symmetry either.

\section{Adding a potential $V(\phi^*\phi)$}
\label{pota}

Consider the action \eqref{GMp} with the addition of a generic potential term respecting the global $U(1)$ symmetry,
\begin{equation}\label{GMV}
 \begin{split}
 S &= \int d^3x \ \Big\{
 -(\partial^t \phi^*)(\partial_t \phi) 
 +A (\partial^i \phi^*)(\partial_i \phi)\\ &
 -B \left[(\partial^i \phi^*)(\partial_i \phi)\right]^2
 -F\, \phi^*\phi\, (\partial^i \partial^j \phi^*)(\partial_i \partial_j \phi)\\ &
 +G\, (\partial^i \phi^*)(\partial_i \phi^*)(\partial^j \phi)(\partial_j \phi)\\ &
 +n f_\kappa(x,\phi^*,\phi)
 - V(\phi^*\phi)
 \Big\}\ .
 \end{split}
\end{equation}
Apart from providing a more generic situation, 
a potential $V(\phi^*\phi)$ is important for a specific reasons:
as shown later, the dispersion relation of the 
phonon becomes in general linear in $q_x$ when a potential $V(\phi^*\phi)$ is considered.

The explicit breaking term $n f_\kappa(x,\phi^*,\phi)$ in \eqref{GMV} still requires $\kappa=k$, 
which in turn solves $\text{EOM}[\varphi]=0$ automatically.
The equation of motion \eqref{onEOM} for the modulus field gets instead modified into
\begin{align}\label{onEOMV}
2 \bar\rho^2 \left\{k^2 \left[A - 2 k^2 \bar\rho^2 \left(B+F-G\right)\right]-V'(\bar\rho^2)\right\} = 0\ .
\end{align}
The energy is given by
\begin{equation}\label{eneV}
 {\cal E}(k,\bar\rho) = k^2 \bar\rho^2 \left[-A + k^2 \bar\rho^2 \left(B+F-G\right) \right] + V(\bar\rho^2)\ ,
\end{equation}
and its minimization with respect to $\bar \rho$ and $k$ returns respectively 
the equation of motion \eqref{onEOMV} and the extra condition 
\begin{equation}\label{dyna}
 2k \bar\rho^2\left[2(B+F-G)k^2 \bar\rho^2 -A\right]= 0\ .
\end{equation}
The composition of the equation of motion \eqref{onEOMV} with the condition \eqref{dyna} 
implies 
\begin{equation}\label{prima}
 V'(\bar\rho^2)=0\ .
\end{equation}

A variation $\delta k$ in \eqref{ansa} corresponds to $\delta \varphi = \delta k\, x$, which does 
not vanish for asymptotically large $|x|$, it even diverges.
In particular, it lies outside of the hypothesis assumed for the variational problem 
to derive the equations of motion. 
One can nevertheless take $\delta k$ variations and derive \eqref{dyna} 
relying on a regularization of the system to a finite box%
\footnote{Comments on how to define Nambu-Goldstone bosons in finite-size systems are given in \cite{Ma:1974tp}.} 
and focusing exclusively on the dynamical character of the symmetry breaking.
Said otherwise, one can neglect the boundary term in the variation of the action
or, equivalently, assume that the boundary conditions be free.
Since the minimization with respect to $\delta \bar \rho$ alone already reproduces the 
only non-trivial equation of motion \eqref{onEOMV}, the study of the energy minimization with respect to both $\delta \bar \rho$ and 
$\delta k$ selects a subset among the solutions to \eqref{onEOMV} (considered with free boundary conditions). 
Requirement \eqref{dyna} is crucial to lead to a flat direction,
 and in turn to a Nambu-Goldstone mode. 

The entries of the matrix $M$ (defined in \eqref{S2}) for the 
quadratic fluctuation action become
\begin{align}
 M_{\sigma\sigma} &= \omega^2 - 2 A k^2 - 4 k^2 \bar\rho^2 (2Fq_x^2 + G q_y^2) \\ \nonumber 
 &\qquad - F \bar\rho^2 (q_x^2+q_y^2)^2 - 2 \bar\rho^2 V''(\bar\rho^2)\ ,\\ \label{tautau_spo}
 M_{\tau\tau} &= \omega^2 -2 n \kappa^2 -2 A q_x^2 - F \bar\rho ^2 \left(q_x^2+q_y^2\right)^2\ ,\\ 
 M_{\sigma\tau} &= M_{\tau\sigma}^* = -2 i k\, q_x \left[A + 2 F \bar\rho ^2\left(q_x^2+q_y^2\right)\right]\ .
\end{align}
where \eqref{prima} has been already considered, so we are taking fluctuations over 
a background that satisfies the energy minimization condition \eqref{dyna}.

The eigenvalues $e_{1,2}$ of $M$ yield the dispersion relations of two modes through $e_{1,2}=0$.
Let us expand these dispersion relations in low momenta and in small $n$:
\begin{align}\label{HV}
&\omega^2  = 2 A k^2
+2 \bar\rho^2 V''(\bar\rho^2)
+\frac{2A^4 n q_x^2}{(A^2+2H\bar\rho^4 V''(\bar\rho^2))^2}\\ \nonumber &
+4Gk^2 \bar\rho^2 q_y^2 
+\frac{4A\bar\rho^2 q_x^2[A(2F+H)k^2 + 2F \bar\rho^2 V''(\bar\rho^2)]}{A^2+2H\bar\rho^4 V''(\bar\rho^2)} +...
\end{align}
and
\begin{align}\label{NGV}
\omega^2 = &\ 2 n \kappa^2
+\frac{4AH\bar\rho^4 V''(\bar\rho^2) q_x^2}{A^2+2H\bar\rho^4 V''(\bar\rho^2)}   \\ \nonumber &
-\frac{4A^3H\bar\rho^2 n \kappa^2 q_x^2}{[A^2+2H\bar\rho^4 V''(\bar\rho^2)]^2}+...
\end{align}
Where again $\kappa$ and $k$ are given by \eqref{kk} and \eqref{sole2},
we have introduced the positive quantity (see comment below \eqref{sole2})
\begin{equation}
H = B+F-G\ ,
\end{equation}
and we assumed $V''(\bar\rho^2)>0$, in accordance with a positive concavity for the energy \eqref{eneV} 
with respect to $\delta\bar\rho$ variations. 

The potential $V(\phi^*\phi)$ does not alter the overall qualitative picture 
already obtained in Section \ref{expl}; the dispersion relations \eqref{HV} and \eqref{NGV} still describe 
respectively a gapped mode and a pseudo-phonon with an $n$-linear squared mass. 
Note however that $V(\phi^*\phi)$ would make a qualitative difference in the 
purely spontaneous case: if $n=0$, the leading $q_x$ term in the dispersion relation of the phonon 
is either quadratic or quartic depending on $V''(\bar\rho^2)$ being trivial or not.
The phonon dispersion relation is 
\begin{equation}
 \omega = c_{\text{ph}} q_x + O(q_x^2)\ ,
\end{equation}
where the longitudinal speed of propagation $c_{\text{ph}}$ is proportional $V''(\bar\rho^2)$. 
Importantly, one can repeat the analysis of Section \ref{homo} also in the presence of $V(\phi^*\phi)$,
still getting the same equation of state $\epsilon = -p$ obtained in \eqref{EOS}.%
\footnote{In the presence of $V(\phi^*\phi)$, 
one needs to consider the equation of motion \eqref{onEOMV} and the minimization condition \eqref{dyna}.
An analogous result was found in \cite{Donos:2015eew} for modulated holographic phases where, 
despite the presence spontaneous modulations, the equilibrium stress-energy tensor assumes the perfect fluid form.}
Specifically, the on-shell value of the Lagrangian density $L_0$ is affected by the potential, 
but Equations \eqref{ene_den}, \eqref{Txx} and \eqref{Tyy} are still valid.

\section{Comments on lattice toy models}
\label{WC}

We still consider a setup with two spatial directions, $x$ and $y$,
and construct vacua with two coexisting space-dependent configurations
(each one similar to that of Section \ref{CDWlike})
characterized by two vectors $\vec k_1$ and $\vec k_2$.
To this purpose we need two independent scalar fields, $\phi_1$ and $\phi_2$. 
We consider doubling the model \eqref{GMV} and adding a cross-term whose role 
is to choose vacua where $\vec k_1$ and $\vec k_2$ form a specific angle $\theta$; namely  
$\vec k_1\cdot \vec k_2 = |\vec k_1||\vec k_2|\cos(\alpha)$ and we want to realize $\alpha = \theta$ dynamically.
Calling $L[\phi^*,\phi]$ the Lagrangian in \eqref{GMV} with $n=0$, we consider
\begin{align}\label{angle}
 &S_\theta = \ \int d^3x \, \Big\{ 
 L[\phi_1^*,\phi_1] + L[\phi_2^*,\phi_2]\\ \nonumber
 -\lambda \ &\Big|\partial_i \phi_1^*\partial_j \phi_2^*\Big(\partial^i \phi_2\partial^j \phi_1
 -\cos^2(\theta)\partial^i \phi_1\partial^j \phi_2 \Big) \Big|^2
 \Big\}\ .
\end{align}
The term in $\lambda$ leads in fact to the minimization of 
\begin{align}\label{mima}
 &(\vec k_1 \cdot \vec k_2)^2 - |\vec k_1|^2 |\vec k_2|^2 \cos^2(\theta)\\ \nonumber & \qquad
 =  |\vec k_1|^2 |\vec k_2|^2 \left[ \cos^2(\alpha) -  \cos^2(\theta) \right]\ .
\end{align}

For field configurations where $\vec k_1$ and $\vec k_2$ form an angle $\theta$, the term in $\lambda$ vanishes 
and the energy of the configuration is the same as that of two non-interacting space-dependent condensates (\emph{i.e.} $\lambda=0$).
Since the $\lambda$ term is the square of a quantity that vanishes on the background, also at the level of linear fluctuations
there are no effects due to $\lambda$. In particular, the stability analysis of the background 
is not affected by $\lambda$.

Model \eqref{angle} enjoys a global $U(1)\times U(1)$ symmetry and leads to vacua where $|\vec k_1|=|\vec k_2|$.
It can however be generalized to cases with a smaller symmetry. As far as the $\lambda$ term is concerned, 
there is an interesting alternative to \eqref{angle},
\begin{equation}\label{nh}
 - \lambda\ \Big|\partial_i \phi_1^*\partial_j \phi_2\Big(\partial^i \phi_2^*\partial^j \phi_1
  -\cos^2(\theta)\partial^i \phi_1^*\partial^j \phi_2 \Big)\Big|^2\ ,
\end{equation}
which preserves only a global $U(1)$ symmetry and still leads to the minimization of \eqref{mima}. 
The phases of the two fields are locked and it is no longer true that 
a global internal symmetry transformation can compensate for a generic translation $\Delta \vec x$. 
This is a way to partially breaking the homogeneity 
down to transformations satisfying
\begin{equation}
 \vec k_1 \cdot \Delta \vec x = \vec k_2 \cdot \Delta \vec x = -\Delta \varphi\ . 
\end{equation}

\section{Discussion and future directions}

The models studied in this paper can be generalized to spacetimes with higher dimensionality.
The addition of transverse spatial directions does not affect the essential aspects of 
the computations and the features of the low-energy modes.%
\footnote{The models could actually be considered also in $1+1$ dimensions, suppressing the $y$ direction.
Still, the essential computational points about the pseudo Nambu-Goldstone modes and propagation along $x$ would 
remain unaltered. On the interpretation level there are however extra subtleties: spontaneous symmetry breaking 
in $1+1$ dimensions is impeded by Coleman-Mermin-Wagner-Hohenberg theorem.
Such obstruction is avoided by quantum field theories in the large $N$ limit,
roughly because the fluctuations which would spoil the condensation of the order parameter are suppressed.
It would be interesting to consider whether, already in $1+1$ dimensions, the models studied here could be regarded 
as effective descriptions of theories in the strict large $N$ limit. Two relevant references on these points are: 
\cite{Argurio:2016xih} where the details of symmetry breaking in (large $N$) holographic field theories in $1+1$ dimensions have been analyzed; 
and \cite{Ma:1974tp} which argues on the existence of Nambu-Goldstone bosons in $1+1$ dimensions, even without a strict 
spontaneous symmetry breaking.}

\subsection{Need for higher derivatives}

We pursued the spontaneous breaking of translation symmetry
following a ``Mexican hat'' strategy applied to spatial gradients,
which implies that quartic terms in the spatial gradients are a necessary ingredient.
This is apparent already from the equation of motion \eqref{onEOM} where,
upon setting $B=F=G=0$, only trivial solutions with either $k=0$ or $\bar\rho=0$
remain. 
Studying the fluctuations about a background \eqref{ansa} with $n=0$ and $V(\phi^*\phi)=0$,
the Nambu-Goldstone frequency does not feature independent longitudinal and 
transverse propagation unless the coupling $F$ is non-trivial, see \eqref{dispe_massless}. 

To the purpose of finding the simplest possible models, 
we restrained the attention to terms 
whose order in spatial derivatives equals the order in the fields.%
\footnote{In \cite{Nicolis:2015sra} a similar claim is made 
in relation to generic effective field theories for phonons.
Comments on power-counting schemes for theories 
enjoying shift symmetries are given in \cite{Son:2005rv}.}
There are three observations about this:
(i) it is remarkable that in order to obtain a propagating 
phonon the Lagrangian in \eqref{GM} needs to be already quite complicated;
(ii) it would be interesting to repeat the present analysis allowing for all the possible consistent terms.%
\footnote{In model \eqref{GM}, the order $4$ for spatial derivatives and $2$ for temporal ones
can be argued by assuming $z=2$ non-relativistic scaling for the time coordinate 
and time-reversal invariance.}
(iii) Higher spatial derivatives are analogous to \emph{frustration}.%
\footnote{In discretized theories a modulated vacuum
arises typically because of the competition between different couplings,
\emph{e.g.} the next-neighbor coupling competing with the 
next-to-next-neighbor coupling. 
If one discretizes the model \eqref{GM}, something similar 
would happen, for instance the next-to-next-neighbor coupling due to the $F$ term competes 
with the next-neighbor coupling due to the $A$ term.}

The $F$ term in \eqref{GM} features more than one derivative applied on the same field.
This could not be avoided even considering spatial partial integrations.
Such a term, if covariantized, would lead to multiple time derivatives applied on the same field,
which in turn would produce Ostrogradsky instabilities \cite{Ostrogradsky:1850fid,Woodard:2015zca}.
As a consequence, the models studied in this paper do not admit a trivial relativistic generalization.

The possibility of breaking translations spontaneously in relativistic theories is 
interesting in relation to the systematic approach of \cite{Nicolis:2015sra} which classifies 
all theoretically possible condensed matter systems in terms of their \emph{spontaneous} 
breaking of Poincar\'e invariance.%
\footnote{The generation of spatially modulated vacua in relativistic field theories has been considered in \cite{Nitta:2017mgk}.}
The models of the present paper could possibly be thought of as effective low-energy descriptions
of relativistic UV theories, assuming that a spontaneous Lorentz symmetry breaking has occurred 
along the renormalization group flow at a scale above the cutoff of the effective theory. 
Studying the possibility of such completions is a future direction.
The embedding of the present models into a systematic effective field theory framework
is a future perspective too. To this regard one could, or perhaps should, consider 
topological terms in the action and couplings with generic dependence on $\phi$.

\subsection{The role of boundary conditions}

In the model \eqref{GMV}, which features a potential $V(\phi^* \phi)$,
it proved to be essential to minimize the energy with respect to both the parameters 
of the ansatz \eqref{ansa}, $k$ and $\bar \rho$. This amounts to minimizing the energy without fixing 
the boundary conditions for the field $\phi$. It is important to recall that the apparent 
larger freedom implied by relaxing the boundary conditions is actually compensated by an
additional requirement on the solutions:
the extra condition descending 
from the energy minimization with respect to variations of $k$.

In order to stress the role played by the boundary conditions we can consider a 
somewhat complementary example: a \emph{kinetic} symmetry breaking where 
the translations are actually broken by boundary conditions.%
\footnote{See \cite{Rabinovici:2007hz} which contains a similar discussion in relation to 
kink solutions induced by boundary conditions.} Let us take the simple example 
of a free real scalar field whose Lagrangian is just $\partial_\mu \psi \partial^\mu \psi$. 
This scalar field can be intuitively related to the phase field $\varphi$ in \eqref{ansa} and in fact one can 
consider an ansatz $\psi = kx$. The equation of motion is satisfied for any value of $k$.
To avoid problems with diverging fields, one can regularize the space to a finite box 
and impose boundary conditions that are compatible with the ansatz for a specific value of $k$.
Said otherwise, if the boundary conditions are fixed, they dictate the value of $k$ regardless
of energy considerations.%
\footnote{If instead the boundary conditions are free, the value of $k$ is arbitrary.}
There is a translation symmetry breaking, but it is forced by 
a ``kinetic" constraint instead of being generated dynamically.%
\footnote{The simplicity of the free real scalar $\psi$ just described is deceitful:
being quadratic in the fields, the model presents a gapless excitation even when the breaking is kinetic, such
gapless mode however is not present in more general theories that feature kinetic translation breaking; 
in particular this gapless mode is not a Nambu-Goldstone because the breaking is not dynamical. 
To stress this point, one can consider again the complex 
field model \eqref{GMV} without letting the boundary conditions be free; $k$ would be related to $\bar\rho$
and the mode which corresponded to the phonon would get a mass proportional to $V'(\bar\rho^2)$.}

\subsection{Topologically non-trivial configurations}

The gradient Mexican hat potential
discloses various possibilities to construct non-trivial topological objects. 
We briefly comment some instances, which are however regarded as future prospects.

A possible topologically non-trivial configuration is the \emph{gradient kink}. 
When the ``potential'' for gradients allows for degenerate absolute minima (\emph{e.g.}  
the solutions \eqref{sole1}), there are 
sectors where the solutions feature a jump in $k$ necessary to connect two degenerate minima.
If the direction of the gradient is along the direction of the step profile,
the gradient kink is \emph{longitudinal} and can be mapped to a normal kink
by redefining the gradient field as the fundamental field, $\partial_x \phi \rightarrow \psi$.

Another slightly more exotic possibility is to combine the topological non-triviality in the 
gradient sector and that resulting from possible degenerate minima in the potential $V(\phi^*\phi)$.

As a final remark, the structure of the the models studied in the present paper,
and specifically the fact that they contain terms with different signs 
and different scaling properties under spatial dilatations,
makes them avoid Derrick's no-go theorem for the existence 
of finite energy solitons \cite{TopSol}.

\subsection{Comments on phenomenology: transport, sound and helical orderings}
\label{sound}

The models studied here can be embedded in larger theories and provide sub-sectors where translations are broken spontaneously
or pseudo-spontaneously. The phenomenological properties of the larger system,
like for example transport, would however depend on the coupling of the translation breaking 
sub-sector with the larger system and, of course, on the characteristics of the latter itself. 

One cannot study the compressional sound mode from the equation of state \eqref{EOS} 
because neither $\epsilon$ nor $p$ depend on the volume.%
\footnote{An analogous equation of state has been studied in \cite{LUONGO:2014haa} and claimed to be consistent with a vanishing speed of sound.
Quintessence models have similar equations of states too, see \cite{Creminelli:2009mu}.}
The first law of thermodynamics actually coincides with the equation of state \eqref{EOS}.
Remarkably, these comments hold independently of the presence of a potential $V(\phi^*\phi)$, see Section \ref{pota}.
The sound mode is instead described upon interpreting the real field $\rho \sin(\varphi - kx)$ 
whose fluctuation is given by $\tau$ (introduced in \eqref{flut}) as a \emph{displacement} field in a target space.
Since the model contains higher derivatives terms,
one cannot just adopt the standard formulae for the speed of sound in terms of the elastic modulus.
One cannot either adopt hydrodynamic formulae, as the system at hand is at zero temperature and zero density.
Rather, the sound mode is studied by the fluctuation analysis of the quadratic action, as described in previous sections.

As a possible future application,
it would be interesting to gain intuition on the qualitative behavior of holographic low-energy modes 
by means of a purely field theoretical toy-model.%
\footnote{For a recent review on holographic sound modes see \cite{Gushterov:2018spg};
for a description of a holographic model closer to the field theories of the present paper we refer to \cite{Amoretti:2018tzw}.}
This is especially interesting in order to match the finite temperature hydrodynamic modes to those of an appropriate $T=0$ theory.

Eventually, the models at hand are technically similar to helical orderings studied in ferromagnetic systems \cite{PhysRev.152.235}.%
\footnote{Or helical orderings in $^3$He \cite{Brauner:2018xhh}.}
There the frequency of the spin-wave modes has been claimed to depend linearly on the longitudinal momentum 
and quadratically on the transverse momentum \cite{Brauner:2010wm}, as in \eqref{NGV} above.

\section{Acknowledgments}

This work has been funded by the Spanish grants FPA2014-52218-P and
FPA2017-84436-P by Xunta de Galicia\\ (GRC2013-024), by FEDER 
and by the Mar\'ia de Maeztu Unit of Excellence MDM-2016-0692
\vspace{10pt}

Special thanks go to Riccardo Argurio for fundamental feedback and suggestions throughout the development of the project.
\vspace{10pt}

\noindent
I want to acknowledge Cristoph Adam, Andrea Amoretti, Daniel Are\'an, Maximilian Attems, Matteo Bertolini,\\ Lorenzo Calibbi, 
Paolo Creminelli, Davide Forcella, Blaise Gout\'eraux, Carlos Hoyos, 
Nicola Maggiore, Javier Mas, Andrea Mezzalira, Giorgio Musso, Alfonso Ramallo, Anibal Sierra-Garc\'ia and Douglas Wertepny for very useful and interesting discussions.
\vspace{10pt}

\noindent
The paper is dedicated to the memory of Rosetta Cervetto.

\appendix 

\section{Related literature}

Some useful references related to the present paper are:
\begin{itemize}
\item
An analysis of pseudo Nambu-Goldstone bosons based on the study of Ward-Takahashi 
identities has been performed for a $U(1)$ symmetry in a generic quantum field theory in \cite{Argurio:2015wgr},
with also a description of its holographic implementation. Similar analyses in a non-relativistic 
context are performed in \cite{Argurio:2015via,Argurio:2017irz}.
\item
The study of spatial pattern formation is intimately related to the study 
of phase transitions from a translationally invariant fluid to a solid.
We refer to the study of Landau which opened the field
\cite{Landau:1937obd} and to two recent papers relying on similar techniques \cite{BOL1,BOL2}.
The branch of research related to stripe and density wave
formation is huge too, for a review 
paper aiming at a systematic organization in the context of the copper-oxides 
we refer to \cite{interco}.
\item
The problem of counting theorems for Nambu-Goldstone
bosons in general circumstances entailing spacetime symmetries and 
non-relativistic contexts has gathered recent interest, see for instance 
\cite{Low:2001bw,Nielsen:1975hm,Watanabe:2011ec,Watanabe:2012hr,Kapustin:2012cr}.
\item
The spontaneous locking of internal and external symmetries 
as the central mechanism for studying effective field theories through 
coset constructions is described for instance in \cite{Nicolis:2013lma,Nicolis:2015sra}.
\item
The interactions between phonons are 
dictated by the symmetry breaking,
this point has been studied in \cite{Leutwyler:1996er}.
It is noteworthy that \cite{Leutwyler:1996er} claims that phonon self-interaction terms are necessarily present
on the basis of consistency and the current algebra. As shown in the main text, a similar conclusion emerges also from the description of the 
translation symmetry breaking dynamics. A precise comparison among these two claims is left for the future.

\item 
The Q-lattice strategy has been exploited in holography also to address 
time-dependent systems  \cite{Biasi:2017kkn,Carracedo:2016qrf}.
It would be interesting to consider whether suitable modifications  of 
the models described in the present paper could allow one to study time translations 
in analogy to spatial translations. 
\item
For a pedagogic review on effective field theories with phonons 
interacting with electrons see \cite{Polchinski:1992ed}.

\end{itemize}


\begin{thebibliography}{}
\bibitem{Donos:2013eha}
  A.~Donos and J.~P.~Gauntlett,
  JHEP {\bf 1404} (2014) 040
  doi:10.1007/JHEP04(2014)040
  [arXiv:1311.3292 [hep-th]].
  
\bibitem{Andrade:2013gsa}
  T.~Andrade and B.~Withers,
  JHEP {\bf 1405} (2014) 101
  doi:10.1007/JHEP05(2014)101
  [arXiv:1311.5157 [hep-th]].
  
\bibitem{Donos:2014oha}
  A.~Donos, B.~Gout\'eraux and E.~Kiritsis,
  JHEP {\bf 1409} (2014) 038
  doi:10.1007/JHEP09(2014)038
  [arXiv:1406.6351 [hep-th]].

\bibitem{Vegh:2013sk}
  D.~Vegh,
  arXiv:1301.0537 [hep-th].

\bibitem{Davison:2013jba}
  R.~A.~Davison,
  Phys.\ Rev.\ D {\bf 88} (2013) 086003
  doi:10.1103/PhysRevD.88.086003
  [arXiv:1306.5792 [hep-th]].

\bibitem{Amoretti:2014zha}
  A.~Amoretti, A.~Braggio, N.~Maggiore, N.~Magnoli and D.~Musso,
  JHEP {\bf 1409} (2014) 160
  doi:10.1007/JHEP09(2014)160
  [arXiv:1406.4134 [hep-th]].
  
\bibitem{deRham:2011rn}
  C.~de Rham, G.~Gabadadze and A.~J.~Tolley,
  Phys.\ Lett.\ B {\bf 711} (2012) 190
  doi:10.1016/j.physletb.2012.03.081
  [arXiv:1107.3820 [hep-th]].

\bibitem{Baggioli:2014roa}
  M.~Baggioli and O.~Pujol\'as,
  Phys.\ Rev.\ Lett.\  {\bf 114} (2015) no.25,  251602
  doi:10.1103/PhysRevLett.114.251602
  [arXiv:1411.1003 [hep-th]].

  \bibitem{Alberte:2017oqx}
  L.~Alberte, M.~Ammon, A.~Jim\'enez-Alba, M.~Baggioli and O.~Pujol\'as,
  Phys.\ Rev.\ Lett.\  {\bf 120} (2018) no.17,  171602
  doi:10.1103/PhysRevLett.120.171602
  [arXiv:1711.03100 [hep-th]].

\bibitem{Alberte:2018doe}
  L.~Alberte, M.~Baggioli, V.~C.~Castillo and O.~Pujol\'as,
  arXiv:1807.07474 [hep-th].

\bibitem{Andrade:2015iyf}
  T.~Andrade and A.~Krikun,
  JHEP {\bf 1605} (2016) 039
  doi:10.1007/JHEP05(2016)039
  [arXiv:1512.02465 [hep-th]].
  
\bibitem{ArkaniHamed:2003uy}
  N.~Arkani-Hamed, H.~C.~Cheng, M.~A.~Luty and S.~Mukohyama,
  JHEP {\bf 0405} (2004) 074
  doi:10.1088/1126-6708/2004/05/074
  [hep-th/0312099].
  
\bibitem{Donos:2015eew} 
  A.~Donos and J.~P.~Gauntlett,
  JHEP {\bf 1603}, 148 (2016)
  doi:10.1007/JHEP03(2016)148
  [arXiv:1512.06861 [hep-th]].
  
\bibitem{2016arXiv160102884C}
  J.~Carrete, W.~Li, M.~L.~Lindsay, D.~Broido, L.~Gallego and N.~Mingo.
  Mat. Res. Lett. {\bf 4} (2016) 4
  doi:10.1080/21663831.2016.1174163
  [cond-mat.mtrl-sci/1601.02884].

\bibitem{Leutwyler:1996er}
  H.~Leutwyler,
  Helv.\ Phys.\ Acta {\bf 70} (1997) 275
  [hep-ph/9609466].

\bibitem{Low:2001bw}
  I.~Low and A.~V.~Manohar,
  Phys.\ Rev.\ Lett.\  {\bf 88} (2002) 101602
  doi:10.1103/PhysRevLett.88.101602
  [hep-th/0110285].
  
\bibitem{Nicolis:2015sra}
  A.~Nicolis, R.~Penco, F.~Piazza and R.~Rattazzi,
  JHEP {\bf 1506} (2015) 155
  doi:10.1007/JHEP06(2015)155
  [arXiv:1501.03845 [hep-th]].
  
\bibitem{Son:2005rv}
  D.~T.~Son and M.~Wingate,
  Annals Phys.\  {\bf 321} (2006) 197
  doi:10.1016/j.aop.2005.11.001
  [cond-mat/0509786].

\bibitem{Gruner:1988zz}
  G.~Gruner,
  Rev.\ Mod.\ Phys.\  {\bf 60} (1988) 1129.
  doi:10.1103/RevModPhys.60.1129

\bibitem{interco}
  E.~Fradkin, S.~A.~Kivelson and J.~M.~Tranquada,
  Rev.\ Mod.\ Phys.\  {\bf 87} (2015) 2.
  doi:10.1103/RevModPhys.87.457
  
\bibitem{PhysRevB.94.134309}
  X.~Chen, H.~D.~Bansal, S.~Sullivan, D.~Abernathy, A.~Aczel, J.~Zhou, O.~Delaire and L.~Shi,
  Phys. Rev. B {\bf 94} (2016) 13
  doi:10.1103/PhysRevB.94.134309  

\bibitem{MIR}
  N.~E.~Hussey, K.~Takenaga and H.~Takagi,
  Phil. Mag. {\bf 84}, (2004)
  doi:10.1080/14786430410001716944
  [arXiv:0404263 [cond-mat]].
 
\bibitem{Delacretaz:2016ivq}
  L.~V.~Delacr\'etaz, B.~Gout\'eraux, S.~A.~Hartnoll and A.~Karlsson,
  SciPost Phys.\  {\bf 3} (2017) no.3,  025
  doi:10.21468/SciPostPhys.3.3.025
  [arXiv:1612.04381 [cond-mat.str-el]].
  
\bibitem{Delacretaz:2017zxd}
  L.~V.~Delacr\'etaz, B.~Gout\'eraux, S.~A.~Hartnoll and A.~Karlsson,
  Phys.\ Rev.\ B {\bf 96} (2017) no.19,  195128
  doi:10.1103/PhysRevB.96.195128
  [arXiv:1702.05104 [cond-mat.str-el]].
  
\bibitem{PhysRevB.54.7489}
  J.~M.~Tranquada, J.~D.~Axe, N.~Ichikawa, Y.~Nakamura, S.~Uchida and B.~Nachumi
  Phys.\ Rev.\ B {\bf 54} (1996)  10, 
  doi:10.1103/PhysRevB.54.7489
  
\bibitem{2018NatMa}
  Y.~Y.~Peng, R.~Fumagalli, Y.~Ding, M.~Minola, S.~Caprara, D.~Betto, M.~Bluschke, G.~M.~De Luca, K.~Kummer, E.~Lefran\c{c}ois, M.~Saluzzo, H.~Suzuki, M.~Le Tacon, X.~J.~Zhou, N.~B.~Brookes, B.~Keimer, L.~Braicovich, M.~Grilli and G.~Ghiringhelli,
  Nature Materials {\bf 17} (12018)  697-702, 
  doi:10.1038/s41563-018-0108-3
 
\bibitem{2018arXiv180904949A}
  R.~Arpaia, S.~Caprara, R.~Fumagalli, G.~De Vecchi, Y.~Y.~Peng, E.~Andersson, D.~Betto, G.~M.~De Luca, N.~Brookes, F.~Lombardi, M.~Saluzzo, L.~Braicovich, C.~Di Castro, M.~Grilli and G.~Ghiringhelli,
  [arXiv:1809.04949 [cond-mat.supr-con]].
  
       
\bibitem{Amoretti:2016bxs} 
  A.~Amoretti, D.~Are\'an, R.~Argurio, D.~Musso and L.~A.~Pando Zayas,
  JHEP {\bf 1705}, 051 (2017)
  doi:10.1007/JHEP05(2017)051
  [arXiv:1611.09344 [hep-th]].

\bibitem{Amoretti:2017frz}
  A.~Amoretti, D.~Are\'an, B.~Gout\'eraux and D.~Musso,
  Phys.\ Rev.\ D {\bf 97} (2018) no.8,  086017
  doi:10.1103/PhysRevD.97.086017
  [arXiv:1711.06610 [hep-th]].

\bibitem{Amoretti:2017axe}
  A.~Amoretti, D.~Are\'an, B.~Gout\'eraux and D.~Musso,
  Phys.\ Rev.\ Lett.\  {\bf 120} (2018) no.17,  171603
  doi:10.1103/PhysRevLett.120.171603
  [arXiv:1712.07994 [hep-th]].
  
\bibitem{Coleman:1985ki}
  S.~R.~Coleman,
  Nucl.\ Phys.\ B {\bf 262} (1985) 263
   Erratum: [Nucl.\ Phys.\ B {\bf 269} (1986) 744].
  doi:10.1016/0550-3213(85)90286-X, 10.1016/0550-3213(86)90520-1

\bibitem{Ostrogradsky:1850fid}
  M.~Ostrogradsky,
  Mem.\ Acad.\ St.\ Petersbourg {\bf 6} (1850) no.4,  385.
  
\bibitem{Woodard:2015zca}
  R.~P.~Woodard,
  Scholarpedia {\bf 10} (2015) no.8,  32243
  doi:10.4249/scholarpedia.32243
  [arXiv:1506.02210 [hep-th]].

\bibitem{Nitta:2017mgk}
  M.~Nitta, S.~Sasaki and R.~Yokokura,
  arXiv:1706.02938 [hep-th].
 
 \bibitem{Rabinovici:2007hz}
   E.~Rabinovici,
   Lect.\ Notes Phys.\  {\bf 737} (2008) 573
    [Les Houches {\bf 87} (2008) 217]
   [arXiv:0708.1952 [hep-th]].
  
\bibitem{TopSol}
  N.~Manton and P.~Sutcliffe,
  Cambridge University Press (2004).
  
\bibitem{Argurio:2016xih}
  R.~Argurio, G.~Giribet, A.~Marzolla, D.~Naegels and J.~A.~Sierra-Garcia,
  JHEP {\bf 1704} (2017) 007
  doi:10.1007/JHEP04(2017)007
  [arXiv:1612.00771 [hep-th]].

\bibitem{Ma:1974tp}
  S.~k.~Ma and R.~Rajaraman,
  Phys.\ Rev.\ D {\bf 11} (1975) 1701.
  doi:10.1103/PhysRevD.11.1701

\bibitem{Gushterov:2018spg}
  N.~I.~Gushterov, R.~Rodgers and R.~Rodgers,
  arXiv:1807.11327 [hep-th].

\bibitem{Amoretti:2018tzw}
  A.~Amoretti, D.~Are\'an, B.~Gout\'eraux and D.~Musso,
  arXiv:1812.08118 [hep-th].

\bibitem{Argurio:2015wgr}
  R.~Argurio, A.~Marzolla, A.~Mezzalira and D.~Musso,
  JHEP {\bf 1603} (2016) 012
  doi:10.1007/JHEP03(2016)012
  [arXiv:1512.03750 [hep-th]].
  
\bibitem{Argurio:2015via}
  R.~Argurio, A.~Marzolla, A.~Mezzalira and D.~Naegels,
  Phys.\ Rev.\ D {\bf 92} (2015) no.6,  066009
  doi:10.1103/PhysRevD.92.066009
  [arXiv:1507.00211 [hep-th]].
  
\bibitem{Argurio:2017irz}
  R.~Argurio, J.~Hartong, A.~Marzolla and D.~Naegels,
  JHEP {\bf 1802} (2018) 053
  doi:10.1007/JHEP02(2018)053
  [arXiv:1709.08383 [hep-th]].
  
 \bibitem{Landau:1937obd}
   L.~D.~Landau,
   Zh.\ Eksp.\ Teor.\ Fiz.\  {\bf 7} (1937) 19
    [Phys.\ Z.\ Sowjetunion {\bf 11} (1937) 26]
    [Ukr.\ J.\ Phys.\  {\bf 53} (2008) 25].
   
\bibitem{Nielsen:1975hm}
  H.~B.~Nielsen and S.~Chadha,
  Nucl.\ Phys.\ B {\bf 105} (1976) 445.
  doi:10.1016/0550-3213(76)90025-0
  
\bibitem{Watanabe:2011ec}
  H.~Watanabe and T.~Brauner,
  Phys.\ Rev.\ D {\bf 84} (2011) 125013
  doi:10.1103/PhysRevD.84.125013
  [arXiv:1109.6327 [hep-ph]].
  
\bibitem{Watanabe:2012hr}
  H.~Watanabe and H.~Murayama,
  Phys.\ Rev.\ Lett.\  {\bf 108} (2012) 251602
  doi:10.1103/PhysRevLett.108.251602
  [arXiv:1203.0609 [hep-th]].
  
\bibitem{Kapustin:2012cr}
  A.~Kapustin,
  arXiv:1207.0457 [hep-ph].

\bibitem{Nicolis:2013lma}
  A.~Nicolis, R.~Penco and R.~A.~Rosen,
  Phys.\ Rev.\ D {\bf 89} (2014) no.4,  045002
  doi:10.1103/PhysRevD.89.045002
  [arXiv:1307.0517 [hep-th]].
  
\bibitem{LUONGO:2014haa}
  O.~Luongo and H.~Quevedo,
  Int.\ J.\ Mod.\ Phys.\ D {\bf 23} (2014) 1450012.
  doi:10.1142/S0218271814500126
  
\bibitem{Creminelli:2009mu}
  P.~Creminelli, G.~D'Amico, J.~Norena, L.~Senatore and F.~Vernizzi,
  JCAP {\bf 1003} (2010) 027
  doi:10.1088/1475-7516/2010/03/027
  [arXiv:0911.2701 [astro-ph.CO]].
  
  \bibitem{PhysRev.152.235}
  R.~Elliott and R.~Lange,
  Phys.\ Rev.\ {\bf 152} (1966) 1,  235-239
  doi:10.1103/PhysRev.152.235
  
\bibitem{Brauner:2010wm}
  T.~Brauner,
  Symmetry {\bf 2} (2010) 609
  doi:10.3390/sym2020609
  [arXiv:1001.5212 [hep-th]].
  
\bibitem{Brauner:2018xhh}
  T.~Brauner and S.~Moroz,
  arXiv:1806.10441 [cond-mat.supr-con].

\bibitem{Biasi:2017kkn}
  A.~Biasi, P.~Carracedo, J.~Mas, D.~Musso and A.~Serantes,
  JHEP {\bf 1804} (2018) 137
  doi:10.1007/JHEP04(2018)137
  [arXiv:1712.07637 [hep-th]].
  
\bibitem{Carracedo:2016qrf}
  P.~Carracedo, J.~Mas, D.~Musso and A.~Serantes,
  JHEP {\bf 1705} (2017) 141
  doi:10.1007/JHEP05(2017)141
  [arXiv:1612.07701 [hep-th]].
  
\bibitem{Polchinski:1992ed}
  J.~Polchinski,
  In *Boulder 1992, Proceedings, Recent directions in particle theory* 235-274, and Calif. Univ. Santa Barbara - NSF-ITP-92-132 (92,rec.Nov.) 39 p. (220633) Texas Univ. Austin - UTTG-92-20 (92,rec.Nov.) 39 p
  [hep-th/9210046].
 
\bibitem{BOL1}
  D.~Bolmatov, E.~T.~Musaev and K.~Trachenko,
 Nat.\ Sci.\ Rep. {\bf 3} 2794
 doi:10.1038/srep02794.
  
\bibitem{BOL2}
  D.~Bolmatov, D.~Zav'yalov, M.~Zhernenkov, E.~T.~Musaev and Y.~Q.~Cai,
 Ann.\ Phys.\ {\bf 363} 221-242 (2015)
 doi:10.1016/j.aop.2015.09.018.
\end{thebibliography}
\end{document}